\documentclass[letterpaper,twocolumn,10pt]{article}
\usepackage{usenix2019_v3}
\usepackage{cite}
\usepackage{amsmath,amssymb,amsfonts}
\usepackage{algorithmic}
\usepackage{graphicx}
\usepackage[shortcuts]{extdash}
\usepackage{textcomp}
\usepackage{xcolor}
\usepackage{url}[hyphens]

\usepackage{hyperref}
\usepackage{caption}[tableposition=to]

\def\BibTeX{{\rm B\kern-.05em{\sc i\kern-.025em b}\kern-.08em
    T\kern-.1667em\lower.7ex\hbox{E}\kern-.125emX}}
\begin{document}

\title{(How) Do people change their passwords after a breach?}

\author{
{\rm Sruti Bhagavatula}\\
Carnegie Mellon University
\and
{\rm Lujo Bauer}\\
Carnegie Mellon University
 \and
 {\rm Apu Kapadia}\\
Indiana University Bloomington
} 

\date{}

\maketitle

\newcommand{\code}[1]{\texttt{#1}}

\newcommand\secref[1]{Section~\ref{#1}}
\newcommand\secrefs[2]{Section~\ref{#1}--\ref{#2}}
\newcommand\secrefssep[2]{Section~\ref{#1},~\ref{#2}}
\newcommand\appref[1]{Appendix~\ref{#1}}
\begin{abstract}

To protect against misuse of passwords compromised in a breach,
consumers should promptly change affected passwords and any similar
passwords on other accounts. Ideally, affected companies should
strongly encourage this behavior and have mechanisms in place to mitigate harm.
In order to make recommendations to companies
about how to help their users perform these and other security-enhancing actions after breaches, we must first
have some understanding of the current effectiveness of companies' post-breach practices.
To study the effectiveness of password-related breach notifications and practices enforced after a breach,
we examine---based on real-world password data
from 249 participants---whether and how constructively participants changed their
passwords after a breach announcement.

Of the 249 participants, 63 had accounts on breached domains; only 33\% of the 63
changed their passwords and only 13\% (of 63)
did so 
within three months of the announcement. New passwords were on average $1.3\times$ stronger than old passwords
(when comparing $\log_{10}$-transformed strength),
though most were weaker or of equal strength. Concerningly, new passwords were overall 
more similar to participants' other passwords, and participants rarely changed passwords 
on other sites even when these were the same or similar 
to their password on the breached domain. Our results highlight the need for more 
rigorous password-changing requirements
following a breach and more effective breach 
notifications that deliver comprehensive advice.
\end{abstract}


\section{Introduction}
\label{sec:intro}

Password breaches have been on the rise, affecting mainstream companies such as Yahoo! and
gaming sites such as League of Legends and Neopets among others~\cite{pwned}.
Stolen passwords have been largely exposed in insecure forms such as
in plain text or by weak hashes (often unsalted or easily guessed through dictionary attacks) such as MD5 and SHA\=/1 hashes,
leaving users
vulnerable unless they change their passwords on the affected sites~\cite{pwned}.
Additionally, when a company suffers a breach involving passwords, rarely 
are the users affected solely on the compromised
domain~\cite{DBLP:conf/ndss/DasBCBW14}.
Previous work has shown that, on average, a user exactly or partially
reuses their passwords on over 50\% of their accounts~\cite{DBLP:conf/ccs/PearmanTNHBCCEF17, florencio2007large, 
DBLP:conf/ndss/DasBCBW14}.
In such cases, when a person's password on one domain is compromised,
they incur the risk that an attacker will be able to gain access to
their other accounts that use similar or the same passwords.
In order to make informed recommendations to
companies on best risk mitigation practices after a breach, it is instructive to
examine people's current password-changing behavior after breaches.

Prior work has explored problems related to data breaches
and changing passwords, 
e.g., how people comprehend data breaches~\cite{data-breaches, equifax-perceptions},
what factors make them more inclined to take action after
breaches~\cite{data-breaches, equifax-perceptions}, and how people change passwords 
in response to reuse notifications~\cite{golla2018site}.
Researchers found that people were more likely to heed advice about actions after security breaches
based on who was giving the advice and often underestimated the
harm that could be incurred as a result of a compromise~\cite{data-breaches, equifax-perceptions}. 
Related to password changes,
researchers found that very few of their participants in an online
study reported intentions to change passwords after being notified
that their passwords were compromised or reused, including because
they believed in the ``invincibility'' of their passwords~\cite{golla2018site}.
These studies are important to understand how to better inform people
about the impact of data breaches
and to understand people's mental models when it comes to taking
action to protect themselves. 
However, we still lack an understanding of the actual extent---empirically measured---to which actions taken by
companies to inform their users after a breach are effective. 

We make a significant effort towards developing this understanding. 
We analyze longitudinal, real-world password data over two years to understand whether people change their passwords after 
a breach and the quality of these password changes. Specifically, we examine: (1)~whether
people with an account on a breached domain
changed their passwords after the breach and how constructive these
changes were; (2)~the extent to which people changed similar passwords
on domains other than
the breached
domains; and (3)~how password changes related to
breaches compare to \emph{all other} password changes.

Our dataset was collected from the home computers of 
249 participants between Jan.\ 
2017 and Dec.\ 2018
and includes \emph{all} passwords used to log onto
online services.
Of the 249 participants, 63 had accounts on one of the breached
domains we studied and were active in the study
at the time of the breach announcement and for three months after.
We found that only 21 of the 63 participants changed their password
after a breach announcement and
only 15 did so within three months of the announcement.
The majority of these changes were in response to a high-risk breach (i.e., the Yahoo!\ breach).
We also found that only a minority of password changes were to stronger passwords 
and that new and old passwords shared a substring on average almost
half the length of the longer of the two passwords.

Participants who changed passwords on the breached domains had on
average 
30 accounts
with similar passwords. Of the 21 participants who changed
passwords, 14 changed at least one similar password within a month
of changing their password on the breached domain. These 14 changed, on average, only four similar passwords
within that month. 

As a baseline for the quality of password changes, we looked at all password changes made by the 249 participants
over the two-year period. A large fraction (69.6\%) of the password changes resulted
in weaker or equal-strength passwords, and old and new passwords on average
shared a substring 85.1\% the length of the longer of the pair. 
Overall, the properties of password changes on 
breached domains were roughly similar to the properties of the baseline password changes, though on average
resulted in more dissimilar passwords.

Our results suggest that current breach notifications are not
effective, in that most users who are affected do not react
sufficiently to mitigate their risk either on the breached domain
or on others. 
Our results clearly indicate that more should be done---through breach
notifications or other means---to induce users to change  passwords
both on the affected domain and especially on other domains, which
users generally ignore. Similarly, additional means are needed to
educate and encourage users to make their new passwords both strong
and different from their existing passwords. 


\section{Related work}
\label{sec:related}

\subsection{Data breaches and security incidents}
Prior work
has studied how people hear about breaches~\cite{DBLP:conf/chi/DasLDH18},
what people comprehend about data breaches~\cite{data-breaches, equifax-perceptions}, and what makes them take action~\cite{data-breaches, equifax-perceptions}. 
Overall, they found that people are more willing to take action after a breach
depending on their 
perceptions of tangible security benefits~\cite{data-breaches} and 
the source of advice 
about actions~\cite{equifax-perceptions}.
A study about breaches and
consumers found that customers' spending 
at a retailer
fell significantly after the retailer suffered a breach~\cite{janakiraman2018effect}, while another survey
found that only a minority of respondents would stop doing business with a company after
a breach~\cite{ablon2016consumer}. Other work has found that people react
to security incidents involving
accounts on a major social network in a variety of ways, from doing nothing to actively seeking out information~\cite{redmiles2019should}.

Users can be alerted about breaches that affect them not just by the organizations that
suffer breaches, but also by dedicated services like
HaveIBeenPwned~\cite{pwned}, LifeLock~\cite{lifelock},
and Enzoic~\cite{enzoic}.
Additionally, password managers 
such as LastPass~\cite{lastpass} and the password manager built into Firefox~\cite{firefox-manager}
alert users if their 
logins are found in data breaches. 
Researchers recently created
a privacy-preserving protocol by which clients can
query breach repositories without revealing the actual 
credentials being queried~\cite{breach-alerting-19}.

\subsection{Password-related behaviors}
Several large-scale password studies have shown that password reuse is rampant~\cite{DBLP:conf/ccs/PearmanTNHBCCEF17, florencio2007large, wash2017can, DBLP:conf/ndss/DasBCBW14}, 
finding that on average people reused over half their passwords~\cite{DBLP:conf/ccs/PearmanTNHBCCEF17, 
DBLP:conf/ndss/DasBCBW14}. Other work showed that people have trouble managing their passwords
and using password managers~\cite{pwd-managers:soups2019}, which
contributes to password reuse~\cite{stobert:pwd-life-cycle}.
Recent work surveyed people's reactions to notifications
that their password
was compromised or was being 
reused on other sites and found that, when advised or required to
change their passwords,
less than a third of respondents reported any intention to
comply~\cite{golla2018site}.
Another study about defenses against
credential stuffing (when an attacker
uses lists of breached usernames and passwords to gain access on a large scale to several other websites) found that when participants were notified
about credential breaches through a privacy-preserving breach 
querying protocol, 
26\% of the notifications caused participants to create passwords that were
at least as strong as their previous
ones~\cite{breach-alerting-19}.

Researchers have measured password-related behaviors in a variety of ways, e.g., by asking participants to install 
password-logging tools~\cite{wash2017can, florencio2007large} and analyzing breached passwords from publicly posted
lists~\cite{DBLP:conf/ndss/DasBCBW14, abbott2018factors} or privately collected datasets~\cite{mazurek2013measuring}.
We leverage data collected through the Security Behavior Observatory
(SBO) (see \secref{sec:data_intro}),
which captures detailed, real-world behavior of home computer users by
instrumenting their operating systems and web browsers~\cite{DBLP:conf/ccs/PearmanTNHBCCEF17, DBLP:conf/hotsos/ForgetKACCT14,
forget2016or}.

\section{Data collection and dataset}
\label{sec:data_intro}


\subsection{Data collection}
We obtained data collected as part of the Security Behavior
Observatory (SBO) project. The SBO is a data-collection infrastructure for a longitudinal study of the security
behaviors of Windows computer users~\cite{DBLP:conf/hotsos/ForgetKACCT14,
forget2016or, DBLP:conf/ccs/PearmanTNHBCCEF17}
that started data collection in October 2014 and ended in July 2019. 
The collected data
includes information about system configuration, system events,
operating system updates,
installed software, and 
browser-related data such
as browsing history, settings, and the presence of browser extensions.
To collect this information, participants' home computers were instrumented with
software that collects data
via system-level processes and
browser extensions.
Specifically, the browser extensions were installed only in
participants' Google Chrome and Mozilla Firefox browsers, 
and recorded every entry into an HTML input field at the time of
browser events such as clicks, key presses, form submissions, and page
loads. The SBO data collection and analysis (including this project) was approved by its institution's ethics review board.

The data analyzed in our study was collected from January 2017 to
December 2018 and includes $249$ participants who participated in the
SBO study for at least 90 days during that period.
Each
participant was enrolled in the SBO 
study at different points in time and for different durations.
The dataset we examine includes information about every entry made into
a password field in a web page, as determined
by the
browser extension, including: a salted one-way hash of the password;
the URL of the form in which the password was
submitted; the strength of the password (represented as
the approximate number of guesses a sophisticated attacker would need to
guess that
password~\cite{DBLP:conf/usenix/MelicherUKBCC17}); and 
hashes of all three-character-or-longer substrings of each password.
Substring hashes are particularly useful for analyses related to partial password reuse, e.g., as
used by Pearman et al.~\cite{DBLP:conf/ccs/PearmanTNHBCCEF17}.
Password guess numbers less than 10 are rounded to 10 for easier comparison
when $\log_{10}$-transformed. Throughout this paper, we represent password strength by its $\log_{10}$-transform (see \secref{sec:results}).

We further filter this raw data as described below.


\subsection{Filtering passwords}
The SBO browser extension collected every entry made into an HTML password field. 
This captured both the entry of correct passwords as well as attempted
logins that failed because an incorrect password was entered.  
The recorded passwords may 
occasionally have been entered by other users on the participant's
computer. A single participant could also have multiple accounts and
passwords on the same domain.

We needed to eliminate any failed login attempts from this dataset
and any 
passwords that did not belong to the participant's main account. We combined collected password
entries across multiple browsers on each participant's machine
and extracted the ``correct'' passwords for a participant by
applying heuristics inspired by
Pearman et al.~\cite{DBLP:conf/ccs/PearmanTNHBCCEF17} and Wash et 
al.~\cite{Wash:2016:UPC:3235895.3235911}, as follows.

We first compiled all password entries on each domain in chronological order.
For each domain, starting from the participant's first password entry on that domain in our dataset, we 
divided the entries into clusters 
where the 
differences between timestamps within one cluster was less than $15$ minutes. We considered the last
entry in this ordinal cluster to be the ``correct'' password of a cluster, i.e., signaling that
the user probably logged in correctly and will not attempt to log into that domain 
again for a while. We then further filtered these clusters 
to remove occasional non-participant logins and each participant's
secondary accounts,
if they had
multiple accounts. If the ``correct'' password of a cluster reappeared in a later cluster, we assumed
that the passwords entered between the two occurrences could have been due to intermittent logins either not by the main user or for less-used accounts.
We only did not consider the entires to be due to intermittent logins when any of the passwords
entered between the two occurrences occurred more frequently than the re-appearing password for the participant or if the password
was submitted over more days in the case of frequency ties.
We do not consider the re-occurrence of an older password to mean the participant changed their password back to an old password
since domains typically do not allow users to change their password to a previously used password.

This process left us with a set of ``correct"
password entries, which is the final dataset we use for
password-related analyses.


\section{Methodology}
\label{sec:methodology}

We study how participants changed their passwords in response to nine data breaches that became public in 2017 and 2018.
We select these breaches based on two broad criteria.

We started with a list of breaches comprised of:
\begin{itemize}
        \item
        Identity Force's list of biggest breaches in 2017~\cite{2017DataBreaches}
	and Digital Information World's list of biggest breaches in 2018~\cite{2018-hacks}; and
	\item breached domains listed on
          \code{haveibeenpwned.com}
          (HIBP) for which
          breached data included passwords~\cite{pwned}. 
          HIBP is a website that keeps track
          of sites that have been compromised and a service that people can query
          to find out whether their personal data has been compromised in a breach.
\end{itemize}

We then selected only those breaches that met the following criteria:
\begin{enumerate}
\item The breach \emph{announcement} date overlapped with the time interval for which we had SBO password data.
\item At least one participant in our dataset entered a password on the breached domain before the breach announcement and remained active
  in the study for 90 days afterward.
\end{enumerate}

This yielded the following nine
breached domains, for which we studied participants' password-change
behavior: Imgur (breach announced Nov.\ 2017)~\cite{imgur_date}, Deloitte
(Sep.\ 2017)~\cite{deloitte_date}, Disqus (Oct.\ 2017)~\cite{disqus_date}, and Yahoo!\ (Feb.\ and Oct.\ 2017)~\cite{yahoo_date_1, yahoo_date_2}, 
MyFitnessPal (Mar.\ 2018)~\cite{myfitnesspal}, Chegg (Sep.\ 2018)~\cite{chegg}, CashCrate (Jun.\ 2017)~\cite{cashcrate}, FLVS
(Mar.\ 2018)~\cite{flvs}, and Ancestry (Dec.\ 2017)~\cite{ancestry}.

For each of these breaches, we first identified participants who
entered passwords on one of these domains, implying that they had an
account on the domain and therefore were \emph{potentially}
affected. We identified these participants as those who entered a
password on at least one of the breached domains before the breach
announcement date and were active in the study for at least 90 days 
after the announcement.
We then checked whether identified
participants changed their password on the
affected domain. If they did,
we checked whether the new password was stronger than the old one,
how similar the new and old passwords were,
whether they also changed similar passwords on other sites, and
whether the password change caused less reuse between the password on their breached
account and other passwords. We next describe the process of
identifying password changes.

\subsection{Identifying password changes}
For each participant who had an account on at least one breached domain, 
we extracted the last password that they 
entered on the domain before the breach
announcement date.
We then looked for the first new password (i.e., different from the last one entered before
the breach announcement) successfully entered on the breached domain 
after the breach announcement. If no new password was found, we
 concluded that the participant had not changed their password.

We also identified whether participants who changed their passwords on
the breached domains changed any similar passwords 
on other domains. We consider two passwords \emph{similar} if
they share a substring that is at least as long as half the length of
the longer password. For example,
the passwords ``iluvDONUTS90'' and ''ih8DONUTS90'' are similar since they share the substring
``DONUTS90'' that is at least half as long as the longer password, ``iluvDONUTS90''. 
We measure similarity by examining passwords similar to the last
passwords entered on any domain before the breach announcement.
If a participant changed their password on a
breached domain, we examine whether they changed any of their similar
passwords in the month that followed.

Even though our dataset directly captures passwords only when they are entered on
participants' home computers, we are able to capture \emph{password
  changes made from other devices too}, because we observe the new
(or unchanged, if they haven't been changed) passwords on the next
login from participants' home computers. Many sites cache authentication credentials and do not
  require users to type in their password on every login. However, we study
  people's behavior over a long enough period that authentication
  credentials, if properly implemented, would have timed out and
  participants would have had to eventually 
  use their passwords to log in.

\subsection{Measuring the effect of password changes}
When participants changed their passwords on a breached domain, we
computed how much stronger (or weaker) the new passwords were (as
described in \secref{sec:data_intro}), the similarity
between their old and new passwords,
and whether the new password was more unique compared to passwords
used on other accounts.

We computed the similarity between old and new
passwords using a normalized similarity metric: the
length of the longest common substring (of length $\geq 3$) between two passwords divided
by the length of the longer password. If two passwords do not share a
substring longer than two characters, we consider them completely dissimilar~\cite{DBLP:conf/ccs/PearmanTNHBCCEF17}.

To examine the relative uniqueness of the old and new passwords, we
computed
the difference in
the amount of (exact or partial) reuse among a participant's passwords before and after
they changed their password on the breached domain 
(described in \secref{sec:results}).
We calculated the extent of reuse of the old password
at the time of the latest entry of the old password, and
the extent of reuse of the new password
a month after the password change,
i.e., a month after the 
 first entry of
the new password on the breached domain. We calculated this reuse after a month to allow time
for the similar passwords on other domains to be changed.
If a participant
changed passwords on more than one breached domain, we computed
the average.

\emph{Computing password reuse:}
To quantify password reuse, we build on the concepts
of
\emph{exact} and \emph{partial} reuse as defined in previous work on password reuse~\cite{DBLP:conf/ccs/PearmanTNHBCCEF17}. 
A password for
a particular account is \emph{exactly} reused if the same participant
uses the same password on another account. A password is
\emph{partially} reused if it shares at least a three-character substring
with another of that participant's passwords~\cite{DBLP:conf/ccs/PearmanTNHBCCEF17}.
An \emph{exactly-or-partially} reused password is one
that satisfies either of these definitions.

Given a password on a domain, we computed its reuse score as the
fraction of that participant's \emph{other} passwords that exactly or partially reuse the password in question. 
We measured reuse based on the latest password entered by 
the participant on each distinct domain before a given point in time.

\subsection{Computing baseline password-change statistics}
\label{sec:baseline_changes}
To provide a baseline against which to compare breach-related password changes,
we computed password-change statistics for all password changes by all 249 participants over
the two years spanned by the dataset. For every instance of a new password per participant---ignoring the first occurrence of
a password since those may have been created prior to the start of data collection---we captured the ratio of the 
strength of the new password to the old. We also computed the length of substrings (of at least 
three characters) shared
by new and old passwords.
Finally, to have a baseline for how strong participants' passwords are overall, we computed the average strength of all of each 
participant's unique passwords entered per domain during
the time period spanned by the dataset, i.e., if a participant
had three unique passwords on \code{google.com} and five on \code{yahoo.com}, we computed the average strength of those eight passwords even
if some of the \code{yahoo.com} passwords were exactly reused on \code{google.com}.


\section{Results}
\label{sec:results}

\subsection{Participants} Of the 249 participants, 60\% identified as female, 39\% as male, and the rest did not provide their gender. 
Ages ranged from 20--81 years with a mean of
34.1. A majority (57\%) were students, and 28\% had professions that involved programming.
Of the 249 participants, 63 had passwords on one or more
of the nine domains involved in a password breach. 
Table~\ref{fig:num_part} shows the number of participants who had an account on each breached domain.

\begin{table}[ht]
\centering
\caption{Number of participants who had an account on each breached domain; some had accounts on more than one of the domains. \label{fig:num_part}}
\begin{tabular}{|l|l|}
\hline
\textit{\textbf{Breached domain}} & \textit{\textbf{Number of participants}} \\ \hline
yahoo.com & 49 \\
myfitnesspal.com & 9 \\
chegg.com & 1 \\
disqus.com & 1 \\
cashcrate.com & 2 \\
flvs.net & 1 \\
ancestry.com & 7 \\
imgur.com & 6 \\
deloitte.com & 1 \\ \hline
Total	& 63 \\ \hline
\end{tabular}
\end{table}

\subsection{Changed passwords}
Only 21 of the 63 participants with passwords on a breached domain changed a password
on the domain after the breach announcement. In total, 23 passwords 
were changed on these domains. Of the 21 participants,
18 were Yahoo!\
users; the remaining 31 Yahoo!\ users (out of 49) did not change their passwords
although all were affected by the breach according to the breach announcement~\cite{yahoo_date_2}.
Two participants changed their 
Yahoo!\ passwords twice, once after each breach announcement.
Two participants changed their password on the breached domain within one month
of the breach announcement, a total of five within two months, and eight
within three months.

\begin{figure}[th]
\centering
\vspace{-3mm}
\includegraphics[height=6cm, width=8cm]{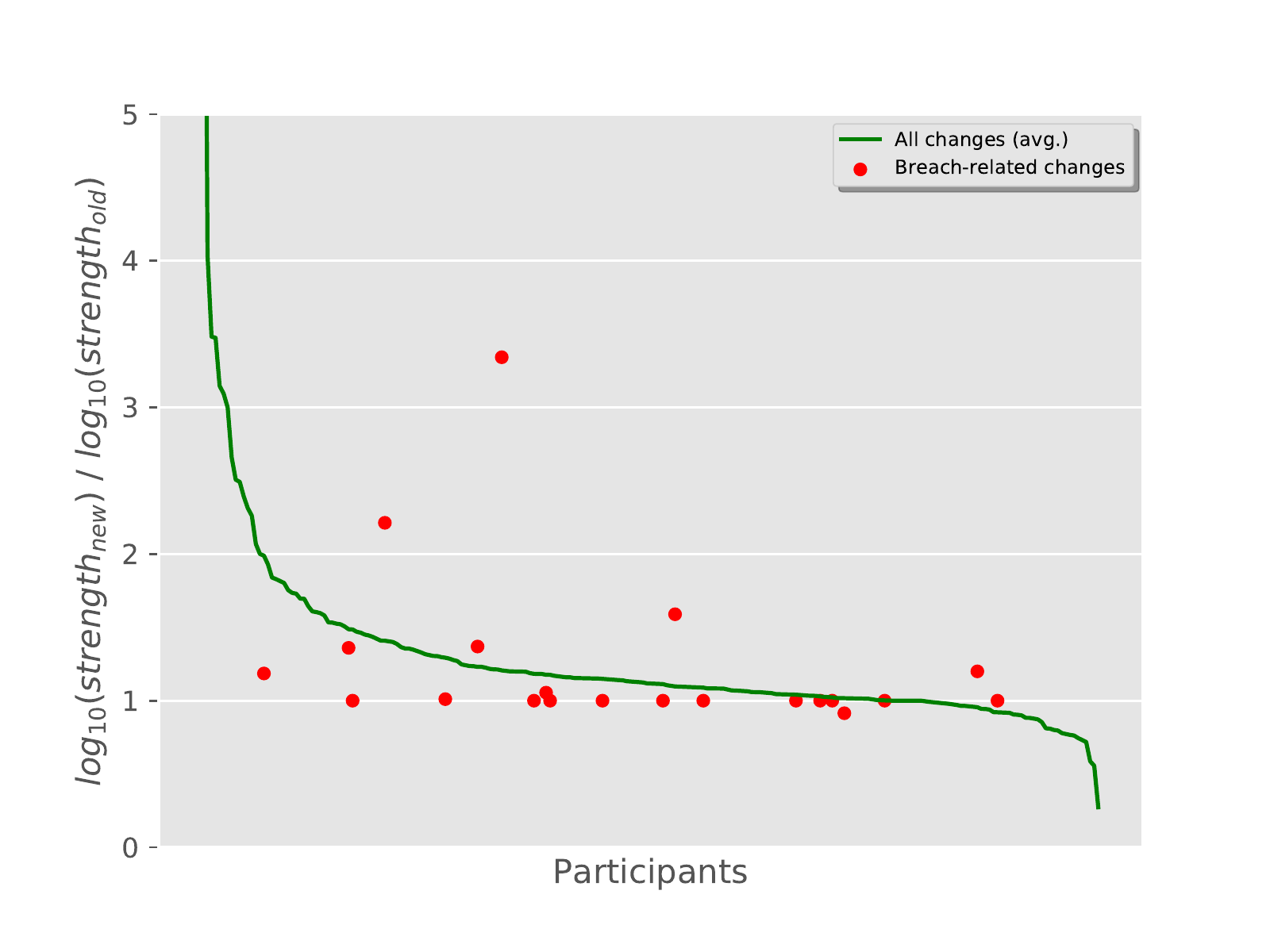}
\caption{Change in password strength across each password change, per
  participant.
  Participants (x axis) are sorted by
  the average amount of improvement in password strength when they change passwords.
 Y-axis values below one indicate that passwords became weaker. \label{fig:strength_quality}}
\end{figure}

\subsection{Quality of new passwords}
For each changed password, we measured the similarity
between the old and the new password,
the strength of the old and the new
passwords, and the extent of password reuse before and after the
password change (see
\secref{sec:methodology}). If a participant changed more than one
password, we report the average results over all the participant's
password changes.

Of the 21 participants who changed their passwords, nine
created stronger (see \secref{sec:data_intro}) passwords and 12 created weaker passwords or ones of equal strength.
On average, participants created new passwords that were $1.3\times$ stronger
than their old passwords after transforming strength on the $\log_{10}$ scale (henceforth, all such comparisons are on $\log_{10}$-transformed strengths).
Seven of the 21 participants who changed their password
created a new password that shared at least a three-character substring
with their old password; for all participants who changed a password, new and old passwords shared a
substring that was on average 41\% as long as the longer of the two
passwords.

The 21 participants who changed a password on a breached domain had, on average, 30
passwords similar to their older breached password (where similar passwords are those that share a substring of
at least half the length of the longer password). 
14 of these participants changed, on average, only 
four of these similar passwords on other sites
within the month after changing their password on the breached site.
These 14 participants changed their similar passwords
to be on average $1.10\times$ stronger than their original password on the breached domain
and $1.18\times$ stronger than the password being changed.
However, the majority (63\%) of the changes resulted in weaker or equal-strength passwords.
Nine participants changed to a password
that shared a substring of three or more characters with their old password;
these nine participants' new passwords on average shared a substring 44\% 
the length of the longer password with their older counterparts.

\emph{Overall, participants changed very few passwords on breached domains and even fewer similar passwords
  on other domains. Even when they did change a password, the change was often not constructive.
}

\subsection{Password reuse}
The passwords changed by our participants were roughly
evenly divided between being less reused and more or equally reused. 
We examined the change in password reuse
for each participant
who changed a password on a breached domain, comparing the reuse before the password change and a month after it. 
For nine participants the new password
on the breached domain
was more reused, for ten it was less reused, and for two it was
equally reused.

\emph{In other words, while participants' new passwords were
slightly stronger and often substantially different from their old
passwords on the same domain, the new passwords on breached sites were still
often similar to
passwords on other domains.}

\subsection{Comparison to baseline password changes}
Looking at \emph{all} password changes by our 249 participants over the two year period,
we observed 223 participants making a total of 3041 password changes, including the changes on the breached domains.
70\% of these password changes resulted in weaker or equally strong passwords.
However, new passwords were on average 
$1.23\times$ stronger than older passwords (again $\log_{10}$-transformed) and the median change in password strength
was neutral (i.e., the old and new passwords were equally strong). All 223 participants
who changed passwords made at least one password change that
involved carrying over a substring of least three characters; in such cases,
old and new passwords shared a substring, on average, 85\% the length of the longer of the two.

\begin{figure}[tb]
\centering
\vspace{-3mm}
\includegraphics[height=6cm, width=8cm]{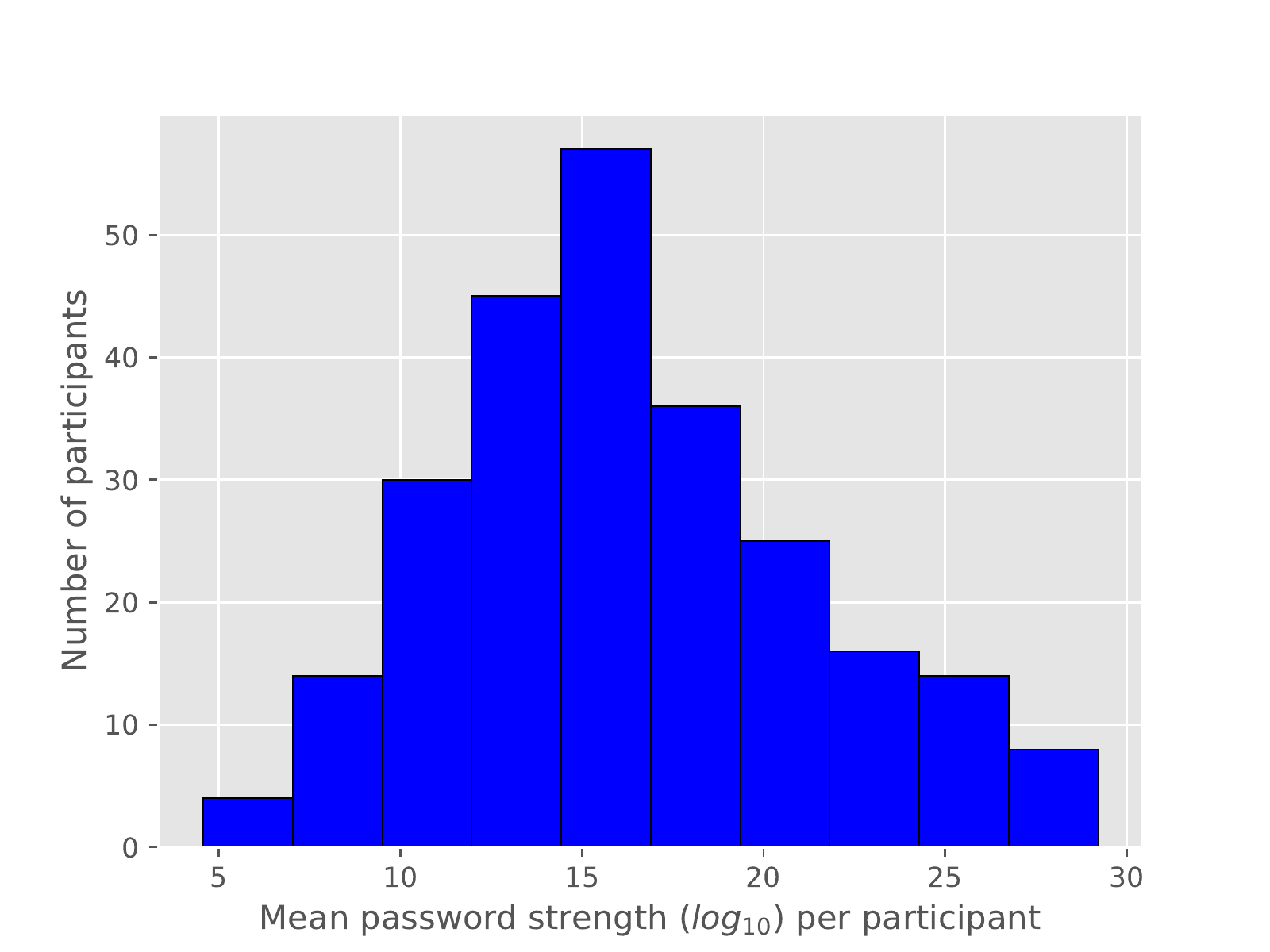}
\caption{The average strength of all of each participant's unique passwords entered per domain.
\label{fig:strength_no_change}}
\end{figure}

Figure~\ref{fig:strength_quality}
shows, per participant, how
changes in password strength for passwords
on breached domains compare to changes in strength of other changed passwords.
The green line on the graph shows the average increase in strength after a password change for
each of the 223 participants over all their password changes.
The red dots show password changes on a breached domain.
Most participants' changes on breached domains resulted in slightly
weaker passwords (red dots above or below the green line) and a minority resulted in substantially stronger
passwords (red dots above the green line), compared to the average changes in password strength.
Figure~\ref{fig:strength_no_change} shows the average strength of all of each participant's unique passwords entered per domain,
computed as described in \secref{sec:baseline_changes}.

\emph{Overall,
password changes showed relatively similar changes in strength, regardless of whether they were on breached domains; 
however, breach-related
password changes resulted in more dissimilar new passwords.}

\section{Limitations}
\label{sec:limitations}

Although our work provides valuable insights
into the effectiveness of post-breach regulations through actions people take after password breaches,
it is subject to a few limitations, including those due to the nature of the data collection.

The participants whose behavior we study are not representative of the
larger population; for example, a quarter had jobs that involve
programming and many were students. Hence, we make no claims with
respect to generalizability.
We also did not have data about the relative importance of each breach
to the data subjects. However,
for the 49 participants with Yahoo!\ accounts, we observed (by examining their web browsing history)
that almost a fourth visited
a Yahoo!\ mail page multiple times a day and another fourth visited such a page
at least once every four days.
This suggests that a large fraction of these
participants were using their Yahoo!\ passwords to protect email
accounts, and hence they should be concerned about the breach.

We do not have data about whether participants were explicitly
notified about a breach; 
rather, we study changes within a window of time after a public breach announcement.

Our analysis of passwords was limited in its precision because
passwords were represented by the hashes of three-character and longer
substrings instead of in plaintext.
This type of information
about passwords has been used previously to study password
reuse~\cite{DBLP:conf/ccs/PearmanTNHBCCEF17} and is
sufficient to reveal substantial reuse in our application.

The data we analyzed was collected from Windows computer users
and limited to passwords entered on Google Chrome and Mozilla
Firefox. Users of non-Windows operating systems
may exhibit behaviors different than the participants in
our dataset. 
Our participants whose password
data we analyzed used Internet Explorer (IE) on average for only 2.86\% of all their browsing
and largely to visit websites that would likely not require them to log in.
Given that IE usage was low among the participants in our dataset and that Windows is the dominant OS for personal
computers~\cite{bott_2013}, 
we do not believe that the unavailability of data about people using non-Windows machines
and of password data from other browsers is likely to fundamentally
affect our findings.

Finally, although their data has been used for several security- or
privacy-related studies~\cite{DBLP:conf/ccs/PearmanTNHBCCEF17,
habib2018away, canfield2017replication, forget2016or},
 the participants enrolled in the SBO study
may be biased towards less privacy- and security-aware
people, given the nature of the SBO data collection
infrastructure. 


\section{Discussion and Conclusions}
\label{sec:discussion}

Out of 63 participants with an account on a breached account, only 21
changed a password on the breached domain, and only eight did so
within three months. Participants on average had 30
passwords similar to their password on the breached domain, but on
average changed only four of these within a month after changing their
password on the breached domain. Even when they changed their password
on a breached domain, most participants changed them to
\emph{weaker} or \emph{equally strong} passwords.
And, regardless of whether participants changed their similar
passwords within a month of the first change, their new passwords on
the breached domains were on average \emph{more} similar to their
remaining passwords.

Some facets of 
good password maintenace
behavior may be difficult for an average user to
grasp~\cite{ur2016users, habib2018away, wash2017can, asgharpour2007mental, hanamsagar2018leveraging}.
For instance, the affinity towards changing to weaker or equal-strength passwords
could be because when people feel
compelled to choose new passwords they have poor awareness of password strength 
or the additional memory burden leads
them to pick weaker passwords~\cite{duggan2012rational, ur2016users}; e.g., they
might change just enough characters to satisfy system requirements. 
Related to partial password reuse, people may find it difficult to understand how their
``different'' password is still similar to other passwords, i.e., they
might be unintentionally partially reusing passwords.  
Potential mitigating efforts could be to integrate
password-reuse trackers within tools that people may already
use and trust to store their passwords. Some password managers, such
as 1Password, already warn users if one of their saved passwords is
reused.  Password managers, including those built into web browsers,
could go further and more actively discourage password reuse.

Overall, our findings suggest that password breach notifications
are failing
dramatically, both at causing users to take action and at causing
users to take \emph{constructive} action.
Regulators should take note of the ineffectiveness or
absence of breach notifications and impose requirements on companies to 
implement better practices~\cite{golla2018site, zou2019beyond, winn2009better,
  zou2019youmight, meters:chi17}.
In particular, they should
encourage companies to send repeat notifications until they have positive
confirmation that the notifications have been understood and that any
instructions have been followed.
Regulators should also require that companies force password resets after a breach
and provide actionable instructions on how to create ``strong''
passwords, describe the risks of password
reuse, and strongly suggest to users that they change passwords beyond the affected
domain. From a preventative standpoint, regulators could incentivize
companies to use an authentication method other than passwords
or to require their users to use two-factor authentication. Companies
should also be required to hash and salt their passwords 
to avoid credential-stuffing and rainbow-table attacks on plaintext or
weakly hashed passwords~\cite{breach-alerting-19,
  oechslin2003making}. 
Regulators could also require services to subscribe to HIBP and to
force users to change their passwords when they encounter a matching
hash. 


\section*{Acknowledgements}
This work was supported in part by the Carnegie Mellon University CyLab Security and Privacy Institute. Parts of the dataset we used were created through work supported by the National Security
Agency under Award No.\ H9823018D0008.  We would also like to thank
Sarah Pearman and Jeremy Thomas for help with understanding and
working with the dataset.

\bibliographystyle{IEEEtranS}
\bibliography{references}

\end{document}